# Predicting MOOCs Dropout Using only two easily obtainable Features from the First Week's Activities

Ahmed Alamri[1], Mohammad Alshehri[1], Alexandra Cristea[1], Filipe D. Pereira[2], Elaine Oliveira[2], Lei Shi[3] and Craig Stewart[4]

[1]Department of Computer Science Durham University, [2] Institute of Computing Federal University of Roraima, [3]Centre for Educational Development University of Liverpool, [4]School of Computing Electronics and Mathematics Coventry University

Corresponding author: Alexandra I. Cristea alexandra.i.cristea@durham.ac.uk

**Abstract.** While Massive Open Online Course (MOOCs) platforms provide knowledge in a new and unique way, the very high number of dropouts is a significant drawback. Several features are considered to contribute towards learner attrition or lack of interest, which may lead to disengagement or total dropout. The jury is still out on which factors are the most appropriate predictors. However, the literature agrees that early prediction is vital to allow for a timely intervention. Whilst feature-rich predictors may have the best chance for high accuracy, they may be unwieldy. This study aims to *predict learner dropout early-on, from the first week*, by comparing several machine-learning approaches, including Random Forest, Adaptive Boost, XGBoost and GradientBoost Classifiers. The results show *promising accuracies (82% - 94%) using as little as 2 features*. We show that the accuracies obtained outperform state of the art approaches, even when the latter deploy several features.

## 1   Introduction

A key concept of MOOCs is to provide open access courses via the Internet that can scale to any number of enrolled students [1]. This vast potential has provided learning opportunities for millions of learners across the world [2]. This potential has engendered the creation of many MOOC providers (such as FutureLearn, Coursera, edX and Udacity)[1], all of which aim to deliver well-designed courses to a mass audience. MOOCs provide many valuable educational resources to learners, who can connect and collaborate with each-other through discussion forums [3]. Despite all their benefits, the rate of non-completion is still over 90% for most MOOCs [4]. Research is still undergoing on whether the low rate of completers indicates a partial failure of MOOCs, or whether the diversity of MOOCs learners may lead to this phenomenon [2]. In the meantime, this problem has attracted more attention from both

---

[1] https://www.mooclab.club/resources/mooclab-report-the-global-mooc-landscape-2017.214/

MOOC providers and researchers, whose goal is to investigate methods for *increasing completion rates*. This starts by determining the *indicators of student dropout*. Previous research has proposed several indicators. Ideally, the earlier the indicator can be employed the sooner the intervention can be planned [5]. Often, combining several indicators can raise the precision and recall of the prediction [6]; however, such data may not always be available. For example, a linguistic analysis of discussion forums showed that they contain valuable indicators for predicting non-completing students [7]. Nevertheless, these features are not applicable to the majority of the student population, as only five to ten percent of the students post comments in MOOC discussion forums [8]. In this paper, we present *a first of its kind research into a novel, light-weight approach based on tracking two* (*accesses to the content pages* and *time spent per access*) *early, fine grained learner activities to predict student non-completion*. Specifically, the machine learning algorithms take into account the first week of student data and thus are able to 'notice' changes in student behaviour over time. It is noteworthy that we apply this analysis on a MOOC platform firmly rooted in pedagogical principles, which has seen comparatively less investigation, namely FutureLearn (www.futurelearn.com). Moreover, we apply our method on a large-scale dataset, which records behaviour of learners in very different courses in terms of disciplines. Thus, the original research question this study attempts to address is:

**RQ**. *Can MOOC dropout be predicted within the first week of a course, based on the learner's number of accesses and time spent per access?*

## 2 Related Research

MOOCs' widespread adoption during their short history, has offered the opportunity for researchers and scientists to study them; with specific focus given to their low rate of completion. This has resulted in the creation of several predictive models that determine student success, with a substantial rise in the literature since 2014 [9].

Predicting students' likelihood to complete (or not to complete) a MOOC course, especially from very early weeks, has been one of the hottest research topics in the area of learning analytics. Kloft et al. [2] used the weekly history of a 12-week-long psychology MOOC course to notice changes in student behaviours over time, proposing a machine learning framework for prediction of dropout and achieving an increase by 15% in prediction accuracy (up to 70% - 85% for some weeks) when compared to baseline methods. However, the model proposed didn't perform correctly during the early weeks of the course. Hong et al. [10] proposed a technique to predict dropouts using learning activity information of learners via applying a two-layer cascading classifier; three different machine learning classifiers - Random Forest (RF), Support Vector Machine (SVM) and Multinomial Logistic Regression (MLR). This study achieved an average of 92% precision and 88% accuracy predicting student dropout. Xing et al. [11], considered active students who were struggling in forums, by designing a prioritising at-risk student temporal modelling approach. This aims to provide systematic insight for instructors to target those learners who are most in need of intervention. Their study illustrates the effectiveness of an ensemble stacking generalisation approach to build more robust and accurate prediction models. As most research on MOOC dropout prediction has measured test accuracy on the same course used for training, this can lead to overly optimistic accuracy estimates. Halawa et al.

[12] designed a dropout predictor using student activity features for timely intervention delivery. The predictor scans student activities for signs of lack of ability or interest, which may lead to long periods of absence or complete dropout. They identified 40% - 50% of dropouts while learners were still active. However, the results provided often failed to specify the precision and recall, or, if they did, they were not detailed at the level of a class (such as for completers and non-completers, separately), but averaged. This is an issue, as it introduces a potential bias, which we further discuss later in this paper.

Additionally, the data is seldom balanced between the classes. This is yet another problem, specifically for MOOCs, where the data distribution between the classes is so skewed (with around 90% of the students belonging to the non-completers class, and only 10% completers). In combination with the averaging of the results, this could lead to over optimistic results. Hence in this paper, we report the results in detail at class level, as well as balancing the data across the classes.

In terms of best performing learning algorithms, the use of random forest (RF) (e.g., [13], [14], [15], [16]) has appeared in the literature among the most frequently used approaches for the student classification tasks. Additionally, Ensemble Methods, such as boosting, error-correcting have been shown to often perform better than single classifiers, such as SVM, KNN and Logistic Regression [17],[18]. In this sense, and to support our early prediction, low feature number approach, we applied the following state-of-the-art classification algorithms to build our model, moving them to the education domain: RF, GradientBoost, AdaBoost and XGBoost. Further improving on the algorithms may render higher accuracy, but is beyond of the scope of this paper.

There have been other studies that have proposed using several machine learning techniques at the same time, to build their prediction models. One study [19] used four different machine learning techniques, including RF, GBM, k-NN and LR, to predict which students are going to get a certificate. However, they used a total of eleven independent variables to build the model and predict the dependent variable – the acquisition of a certificate (true or false); whereas our model uses only two independent variables (the number of accesses and the time spent on a page). Additionally, their results indicated that most learners who dropped out were likely to do so during the first weeks. This supports our assumption that early prediction is possible and can be accurate. Importantly, unlike our approach of using only two independent variables (features/attributes), most prior research used many. For example, [2] employed nineteen features, including those that capture the activity level of learners and technical features. Promisingly our model, despite using only two features from only the first week of each course, can also achieve a 'good enough' performance, as shall be further shown.

## 3 Methodology

### 3.1 Data Preparation

This study has analysed data extracted from 21 runs of 5 FutureLearn-delivered courses from The University of Warwick between 2013 and 2017. The number of accesses and the time spent have been computed for each student. The courses analysed can be roughly classified into 4 main themes: literature (Shakespeare and His World); Psychology (The Mind is Flat) and (Babies in Mind); Computer Science (Big Data) and Business (Supply Chains). Runs represent the number of repeated delivery for each of the five courses. The number of runs for each course is (5, 6, 6, 3 and 2, respectively) whereas the set number of weeks required for studying each course is (10, 6, 4, 9 and 6). In total, they involve the activities of 110,204 learners, who accessed 2,269,805 materials, with an average of around 21 materials accessed per student.

Some courses offer quizzes every week, on subjects of different nature and/or difficulty level, whereas others skip some of the weeks. Due to all the above variations between the courses, we have considered it best to analyse each courses independently, merging only the data from different runs of each course. The latter was made possible, as all courses had runs (within that course) of identical length and similar structure.

In order to determine if there is a normal distribution of variables in each group (completers and non-completers), the Shapiro–Wilk test was used. On determining that distribution was non-parametric, the Wilcoxon signed-rank test was applied, to determine if there is a significant difference between completers and non-completers.

In order to prepare and analyse the data, we next define the employed feature extraction and selection technique, as well as the machine learning algorithms previously identified to address our research question. To begin with, the raw dataset was refined, removing all students who enrolled but never accessed any material. We dealt with those learners separately, based on even earlier parameters (such as the registration date) [20]. Subsequent to this there were 110,204 remaining learners to be studied, of which 94,112 have completed less than 80% and only 16,092 have completed 80% or more of the materials in the course. The reason of selecting 80% completion as a sufficient level of completion (as opposed to, e.g., 100% completion) is based on prior literature and our previous papers [20] [21] [22], where we consider different ways of computing completion. Moreover, the total number of those who completely accessed 100% of the steps was relatively low.

In terms of early prediction, we have opted for the first week, as this methodology is one of the most difficult and least accurate approaches when comparing with the current state of the art in the literature. Alternatively, a relative length (e.g., 1/n days of the total length of each course) could have been used. However, in practice, this tends to use later prediction data than our approach (e.g., $1/4^{th}$ of a course is 1 week for Babies in Mind, but 2.5 weeks for Shakespeare and his Work).

### 3.2    Features Selection

Unlike the current literature, this study determined to minimise the number of indicators utilised. In order to check which indicators are more important, we use an embedded feature selection method that evaluates the importance of each feature by the time that the model is training. As we used tree-based ML algorithms, the metric to measure the importance of each feature was the Gini-index [23]. **Fig. 1** shows the most important features for each course.

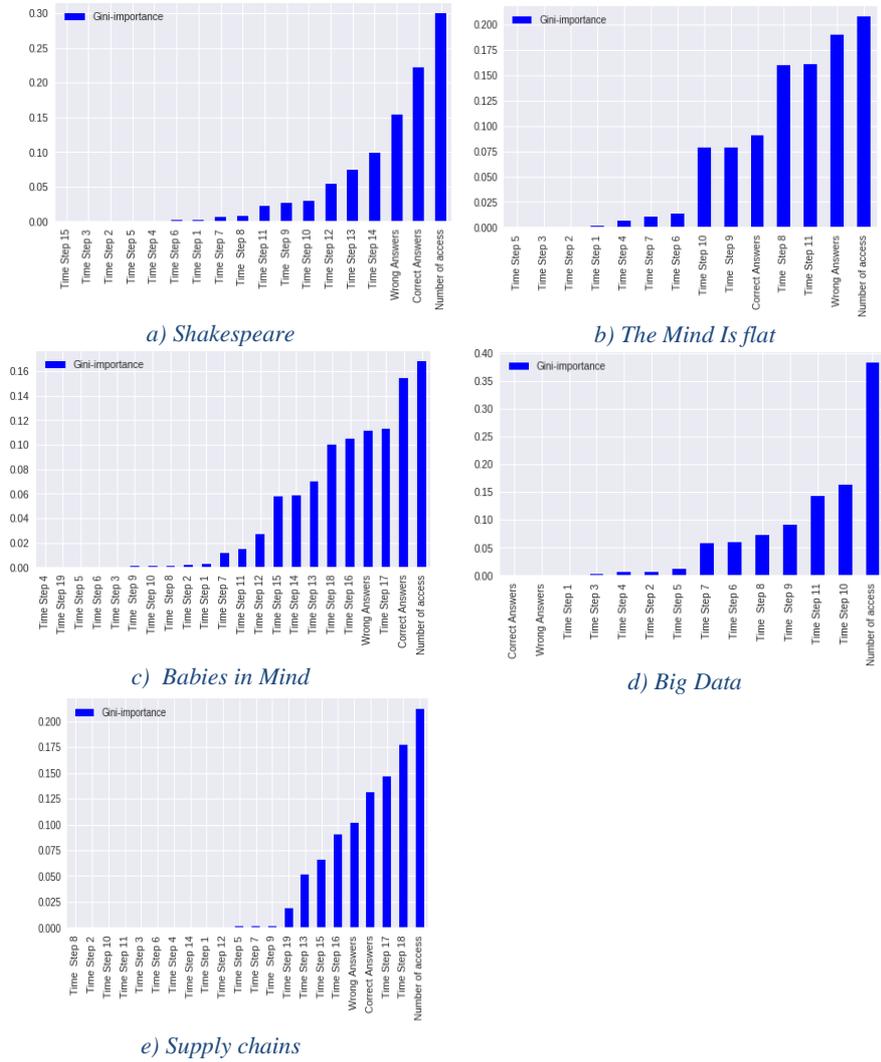

*Fig. 1. Gini-index for the features in the five courses.*

As one of the goals of this study was to create a simple model, we focused on specific features which could be used for various MOOCs – this was done to enhance the generalisation and applicability of the findings for the providers. Therefore, we applied four features to predict the student completion, as follows. ***Number of Accesses*** represents the total number of viewed steps (articles, images, videos), whereas ***Time Spent*** represents the total time spent to complete each step. ***Correct answers*** represents the total number of correct answers and ***Wrong answers*** represents the total number of wrong answers (see **Fig.1** Gini-importance for all the five courses).

We concluded that ***Time spent***, and ***Number of access*** are the most important features, since those two features are not only easy to obtain for most courses, but also results show that ***Time spent*** in each step is playing a critical role to predict the student completion. Moreover, the number of accesses was, in general, an important feature in all the courses. Furthermore, it should be taken in consideration that some courses do not have quizzes in every week; in this case the *Wrong answer* and *Correct answers* features do not play any role to predict the student's completion in those courses (see big data course in **Fig.1** (*d*)).

### 3.3 Building machine learning models

To build our model, we employed several competing ML ensembles methods, as follows: Random Forest (RF) (Breiman, 2001), Gradient Boosting Machine (Gradient Boosting), [24] Adaptive Boosting (AdaBoost) [25] and XGBoost [17] to proceed with exploratory analysis. Ensembles refers to those learning algorithms that fit a model via combining several simpler models and converting weak learners into strong ones [26]. In cases of binary classification (like ours), Gradient Boosting uses a single regression tree to fit on the negative gradient of the binomial deviance loss function [24]. XGBoost, a library for Gradient Boosting, contains a scalable tree boosting algorithm, which is widely used for structured or tabular data, to solve complex classification tasks [17]. Adaboost is another method, performing iterations using a base algorithm. In each interaction, Adaboost uses higher weights for samples misclassified, so that this algorithm focuses more on difficult cases [25]. Random Forest is a method that use a number of decision trees constructed using bootstraping resampling and then applying majority voting or averaging to perform the estimation [27].

After comparing the above methods based on a training and test set division of 70% / 30% respectively, in order to more accurately estimate the capacity of the different methods to generalise to an independent (i.e., unknown) dataset (e.g., to an unknown run of a course), and to avoid overfitting, we have also estimated the prediction accuracy based on 10-fold cross-validation, a widely used technique to evaluate a predictive model [28]. In order to obtain confidence intervals for all the performance metrics (accuracy, precision, recall, F1-score), we have attempted to predict student completion a hundred times, by choosing testing and training sets randomly [29].

### 4 Results

This section details the results of our prediction task of using the first week to determine if the learners selected in the above section are to be completers or non-completers, based on different algorithms. Table 1 compares Random Forest (RF), Adaboost Classifier, XGBoost Classifier and GradientBoosting Classifier methods for all five courses, reporting on some of the most popular indicators of success: accuracy, precision, recall, and the latter two combination, the F1 score.

In general, all algorithms achieved almost the same result, indicating that regardless of the employed model, the features selected in this study proved to be powerful in predicting completers and non-completers. Moreover, our predictive models were able to achieve high performance in each class (completers '1' and non-completers '0') as shown in Table 1. *The prediction accuracy varies between 83%-93%.* We can see that the best performing course, across all four methods applied, is the 'Shakespeare' course.

Table 1. Prediction performance for balanced data (oversampling)

| | Accuracy | [+-] | Precision 0 | [+-] | 1 | [+-] | Recall 0 | [+-] | 1 | [+-] | F1 Score 0 | [+-] | 1 | [+-] |
|---|---|---|---|---|---|---|---|---|---|---|---|---|---|---|
| **Big Data** | | | | | | | | | | | | | | |
| Random Forest | 91.08 | 0.04 | 98 | 0.03 | 85 | 0.07 | 83 | 0.09 | 98 | 0.02 | 90 | 0.05 | 91 | 0.04 |
| Gradient Boosting | **91.43** | **0.04** | 99 | 0.01 | 85 | 0.07 | 83 | 0.09 | 99 | 0.01 | 90 | 0.05 | 92 | **0.04** |
| AdaBoost | 91.37 | 0.04 | 99 | 0.01 | 85 | 0.07 | 82 | 0.08 | 99 | 0.01 | 90 | 0.05 | 92 | 0.04 |
| XGBBoost | **91.38** | **0.05** | 99 | 0.02 | 85 | 0.08 | 82 | 0.09 | 99 | 0.01 | 90 | 0.05 | 92 | **0.05** |
| **The Mind is Flat** | | | | | | | | | | | | | | |
| Random Forest | **87.65** | **0.05** | 98 | 0.04 | 80 | 0.07 | 76 | 0.08 | 98 | 0.03 | 86 | 0.05 | 88 | **0.04** |
| Gradient Boosting | 87.91 | 0.04 | 98 | 0.02 | 80 | 0.06 | 76 | 0.08 | 99 | 0.02 | 86 | 0.05 | 89 | 0.04 |
| AdaBoost | **87.78** | **0.04** | 99 | 0.03 | 80 | 0.07 | 76 | 0.08 | 99 | 0.02 | 86 | 0.05 | 89 | **0.04** |
| XGBBoost | 87.94 | 0.05 | 99 | 0.03 | 80 | 0.06 | 76 | 0.08 | 99 | 0.02 | 86 | 0.05 | 89 | 0.04 |
| **Babies in Mind** | | | | | | | | | | | | | | |
| Random Forest | 82.69 | 0.05 | 96 | 0.04 | 75 | 0.08 | 67 | 0.14 | 97 | 0.03 | 79 | 0.06 | 84 | 0.05 |
| Gradient Boosting | **83.47** | **0.05** | 98 | 0.04 | 75 | 0.08 | 67 | 0.1 | 98 | 0.03 | 80 | 0.07 | 85 | **0.05** |
| AdaBoost | 83.30 | 0.05 | 98 | 0.05 | 75 | 0.08 | 67 | 0.1 | 99 | 0.03 | 80 | 0.07 | 85 | 0.05 |
| XGBBoost | **83.41** | **0.06** | 98 | 0.04 | 75 | 0.08 | 67 | 0.11 | 99 | 0.02 | 80 | 0.08 | 85 | **0.05** |
| **Supply Chain** | | | | | | | | | | | | | | |
| Random Forest | **92.08** | **0.11** | 99 | 0.06 | 86 | 0.17 | 85 | 0.22 | 99 | 0.05 | 91 | 0.13 | 92 | **0.1** |
| Gradient Boosting | 93.40 | 0.1 | 99 | 0.03 | 88 | 0.18 | 86 | 0.2 | 99 | 0.03 | 92 | 0.11 | 93 | 0.1 |
| AdaBoost | **93.11** | **0.1** | 99 | 0.05 | 88 | 0.17 | 86 | 0.19 | 99 | 0.04 | 92 | 0.11 | 93 | **0.1** |
| XGBBoost | 93.14 | 0.09 | 99 | 0.03 | 87 | 0.16 | 86 | 0.19 | 99 | 0.02 | 92 | 0.11 | 93 | 0.09 |
| **Shakespeare** | | | | | | | | | | | | | | |
| Random Forest | 93.03 | 0.09 | 99 | 0.04 | 88 | 0.15 | 86 | 0.18 | 99 | 0.04 | 92 | 0.11 | 93 | 0.09 |
| Gradient Boosting | **93.26** | **0.11** | 99 | 0.04 | 88 | 0.17 | 86 | 0.22 | 99 | 0.03 | 92 | 0.13 | 93 | **0.1** |
| AdaBoost | 93.10 | 0.1 | 99 | 0.06 | 88 | 0.16 | 86 | 0.19 | 99 | 0.05 | 92 | 0.11 | 93 | 0.09 |
| XGBBoost | **93.20** | **0.09** | 99 | 0.05 | 88 | 0.16 | 86 | 0.2 | 99 | 0.05 | 92 | 0.11 | 93 | 0.09 |

0: Non- Completer Group, 1: Completer Group , [+-]: Error of margin over 100 prediction times

The chart below (**Fig.2**) illustrates the median of the time spent by completers and non-completers on the first step of the first week across all the five courses. Results show that completers spent between 66% to 131% more time than non-completers in Big Data and Shakespeare, respectively. Supply Chain recorded the highest ratio between both groups of learners, with 601% more time spent by completers. However, the difference between the two groups was lower, i.e., 25% more for completers, for Babies in Mind.

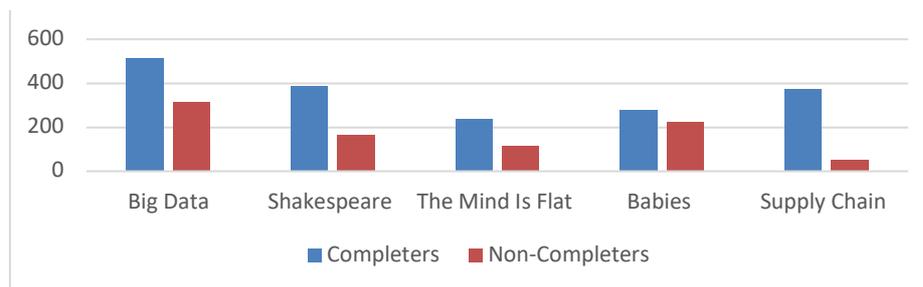

*Fig. 2. Time spent on the first step of the first week (week one) by completers and non-completers.*

Additionally, the Shapiro test was used to determine the normal distribution of variables in each group (completers and non-completers). The results show that the time spent is not normally distributed (p-value < 2.2e-16) in all courses. Therefore, the Wilcoxon test was used to determine if there is a significant difference between the completers and non-completers groups. The results show that two data sets are significantly different in all courses – in other words, that the completers spend not only more time on average than the non-completers, but that this difference is significant.

## 5    Discussion

We have selected four of the most successful methods for classification problems, applying them in the domain of learning analytics in general, and on completion prediction in particular.

Another candidate was SVM, which we did apply, but which was less successful with a linear kernel and would possibly need a non-linear kernel to improve accuracy. In terms of the variation of accuracy, precision, recall and F1 score between courses, the best performing course, 'Shakespeare', was the longest (10 weeks), with a relatively good amount of data available (5 runs). The worst performing course, on the other hand, 'Babies in Mind', was the shortest (4 weeks).

Thus, for all methods, long courses, such as 'Shakespeare' (spanning over 10 weeks) and 'Big Data' (taking 9 weeks), perform better. Moreover, it seems that the longer the course, the better the prediction, as the prediction for the 10-week course on Shakespeare outperforms the prediction of the 9-week course on Big Data across all methods consistently, for both training and test set. *Our accuracy is very high - between 82-94% across all courses. This is equivalent to the current best in breed from the literature, but utilised far fewer indicators to achieve a much earlier success.* This is due to the careful selection process of the two features, which are both generic, as well as informative. One important reason of why the two early, first week features were enough for such good prediction is the fine granularity of the mapping of these features – for each FutureLearn 'step' (or piece of content) we could compute both number of accesses as well as time spent. Thus, the application of the features for the first week transformed into a multitude of pseudo-features, which would explain the increased prediction power. Nevertheless, this method is widely applicable and does not detract from the generalisability of our findings.

Importantly, we have managed to predict only based on the first week of the course, how the outcome will look like. For some courses, this represents prediction based on a quarter of the course (e.g., for Babies in Mind). For others, the prediction is based on data from one tenth of the course, which is a short time to draw conclusions from.

A few further important remarks need considered. Firstly, the data pre-processing is vital: here we want to draw the attention especially to the balancing of the data. For such extremely skewed datasets as encountered when studying MOOC completion, where averages of 10% completion are the norm, prediction can 'cheat' easily: by, e.g., just predicting that all students fail, we would obtain a 90% completion rate! In order to avoid such blatant bias, we balance the data.

Furthermore, the way the data is reported is important. Many studies just report the average for the success measure (be it accuracy, recall, precision, F-score, etc.) over the two categories. As we can see above, the difficulty in the problem we are tackling is the prediction of the completers: thus, it would be easy to hide the poor prediction on this 'hard' category, by averaging the prediction across categories and students. To ensure this is not happening, we provide in this paper separate measures for each category, so the results we are reporting don't suffer from this bias.

## 6     Conclusion

In this paper, we have shown the results from our original study that demonstrates that we can provide reliable, very early (first week) prediction based on two easily obtainable features only, thus via a light-weight approach for prediction, which allows for easy and reliable implementation across various courses from different domains. Such an early and accurate predictive methodology does not yet exist beyond our research and as such this is the first in this class of model. We have shown that these two features can provide a 'good enough' performance, *even outperforming state of the art solutions involving several features*. The advantage of such an approach is clear: it is easier and faster to implement across various MOOC systems, and does require the existence of only a limited amount of information points. The implementation itself is light-weight, and is much more practical when considering an on-the-fly response, and has a limited cost in terms of implementation resources, and more importantly, in terms of time. The results we have obtained are based on balanced datasets, and we report success indicators across both categories, completers and non-completers. We thus avoid both bias in terms of unbalanced datasets, as well as bias based on averaging.

**Acknowledgment.** We would like to thank FAPEAM (Foundation for the State of Amazonas Research), through Edital 009/2017, for partially funding this research.